\documentclass[eqsecnum]{aastex}
\usepackage{bm}
\usepackage{mathrsfs}
\usepackage{natbib}
\usepackage{graphicx}
\usepackage{epstopdf}
\usepackage{amsmath}
\usepackage{amssymb}
\usepackage{amstext}
\usepackage{amsfonts}
\usepackage{color}
\shorttitle{Pointing LISA-like gravitational wave detectors}
\shortauthors{Jani et al.}
\begin{document}
\title{Pointing LISA-like gravitational wave detectors}

\author{Karan Jani\altaffilmark{1,2}, Lee Samuel Finn\altaffilmark{3,1}, Matthew J. Benacquista\altaffilmark{4}}
\altaffiltext{1}{Department of Astronomy and Astrophysics, Pennsylvania State University, University Park, PA 16802}
\altaffiltext{2}{Current address: School of Physics, Georgia Institute of Technology, Atlanta, GA 30332-0430}
\altaffiltext{3}{Department of Physics, Pennsylvania State University, University Park, PA 16802}
\altaffiltext{4}{Center for Gravitational Wave Astronomy, The University of Texas at Brownsville, 80 Fort Brown, Brownsville, TX 78520}

\begin{abstract} 
Space-based gravitational wave detectors based on the Laser Interferometer Space Antenna (LISA) design operate by synthesizing one or more interferometers from fringe velocity measurements generated by changes in the light travel time between three spacecraft in a special set of drag-free heliocentric orbits. These orbits determine the inclination of the synthesized interferometer with respect to the ecliptic plane. Once these spacecraft are placed in their orbits, the orientation of the interferometers at any future time is fixed by Kepler's Laws based on the initial orientation of the spacecraft constellation, which may be freely chosen. Over the course of a full solar orbit, the initial orientation determines a set of locations on the sky were the detector has greatest sensitivity to gravitational waves as well as a set of locations where nulls in the detector response fall. By artful choice of the initial orientation, we can choose to optimize or suppress the antennas sensitivity to sources whose location may be known in advance (e.g., the Galactic Center or globular clusters).
\end{abstract}

\keywords{methods: observational --- gravitational waves --- telescopes --- instrumentation: detectors --- techniques: interferometric --- space vehicles}

\section{Introduction}\label{sec:intro}

In the current effort to open the gravitational wave astronomy frontier, a space-based laser-interferometric gravitational wave detector operating in the 0.1~mHz to 100~mHz waveband is far and away the most promising marriage of technological capability and number and variety of strong sources. In its most studied form a space-based laser interferometric gravitational wave detector involves three free-flying sciencecraft --- acting as the stations of an interferometer --- inhabiting 1~AU circumsolar orbits arranged so that the sciencecraft form an equilateral triangle constellation, inclined to the ecliptic plane by $60\deg$, with arm lengths of several million km \citep{jennrich:2011:nrh}. Once set in their orbits the orientation of the interferometer with respect to the celestial sphere is determined by Kepler's laws of motion; correspondingly, the initial orientation of the equilateral triangle in its plane affects the detector's sensitivity to gravitational waves sources in different regions of the sky. In all present proposals only one of the sciencecraft is equipped to measure the relative velocity to both of the other two sciencecraft; so, only one interferometer can be synthesized \citep{jennrich:2011:nrh,baker:2011:shl,baker:2011:sll,baker:2011:sll:1,baker:2011:sml}. Here we show how the freedom to choose the initial orientation of the sciencecraft constellation --- and, thus, the orientation of the interferometer --- affects the detector's sensitivity to gravitational waves associated with sources in the direction of the Galactic Center.

\citet{decher:1980:dao} first investigated the design aspects of a space-based, laser interferometric gravitational wave detector. The mission and detector they described quickly converged to the modern ``LISA'' concept that, in its most mature form, is described in several  technical assessments performed for ESA \citep{danzmann:2011:luh,jennrich:2011:nrh}.\footnote{``LISA'' was the name adopted for a joint ESA-NASA mission: in the U.S.~as part of NASA's Beyond Einstein Program and in Europe as part of ESA's Cosmic Vision 2015-2025 Program. While that mission was ended when NASA was unable to meet its commitments, missions based on the LISA design but rescaled for different cost-caps are being considered by both ESA and NASA \citep{baker:2011:shl,baker:2011:sll,baker:2011:sll:1,baker:2011:sml,jennrich:2011:nrh}. Here we use the term ``LISA'' to refer to any space-based, laser-interferometric gravitational wave detector whose mission profile is similar to the former joint ESA-NASA mission that was ranked as a large space mission priority in \citep{committee-for-a-decadal-survey-of-astronomy-and-astrophysics:2010:nwn}.} 

In the modern LISA concept each of the three sciencecraft follow independent, circumsolar orbits that lead or lag Earth in its orbit by tens of degrees. Each sciencecraft hosts a freely floating reference mass (a ``gravitational reference sensor'', or GRS) that it simultaneously shields from the thermal and particle space environment and tracks to ensure that the sciencecraft trajectory is as close to a free trajectory through space-time as technology allows. Modulated laser signals passed between the sciencecraft are used to measure the relative velocities of the three reference masses. Gravitational waves passing through the constellation lead to correlated disturbances in these relative velocity measurements. 

The gravitational wave antenna that is synthesized from the relative velocity measurements of the three reference masses behaves like an interferometer with one sciencecraft acting as a beamsplitter and the other two acting as end-stations \citep{armstrong:2003:tdi,dhurandhar:2005:ti}. 
Like an interferometer its greatest sensitivity is to sources in the directions normal to the plane  defined by the three sciencecraft. Also like an interferometer the antenna has four nulls --- directions in which it is entirely insensitive to gravitational wave sources --- that lie in the plane. The direction of these nulls is tied to the orientation of the sciencecraft in their plane. The requirement that the sciencecraft each travel freely through space yet maintain, over several orbital periods, the same relative configuration, strongly constrains the sciencecraft orbits \citep{folkner:1997:los,folkner:2001:los,dhurandhar:2002:aat,dhurandhar:2005:fol}, leading each null to trace-out a ``figure-eight'' pattern on the sky over the course of a year.  Where these patterns appear on the sky is set by the initial orientation of the sciencecraft in their plane. Our purpose here is to demonstrate how the choice of initial constellation orientation affects the sensitivity of LISA concept detectors to the direction hosting the greatest number of gravitational wave sources: i.e., the Galactic Center. 

Section \ref{sec:measure} of this manuscript describes our sensitivity measure for a LISA concept gravitational wave antenna as it orbits Sol. Section \ref{sec:disc} examines the sensitivity of a LISA concept mission to gravitational wave sources in the direction of Milky Way Galactic Center and some of the more important Milky Way globular clusters. We summarize our conclusions in Section\ref{sec:concl}. 

\section{Sensitivity measure}\label{sec:measure}

We base our measure of LISA's sensitivity to point gravitational wave sources on its low-frequency angular response. For our purposes the three LISA sciencecraft correspond to the beamsplitter and end-mirrors of a Michelson delay-line interferometer. Each sciencecraft is on an independent, 1~AU circumsolar orbit. These orbits are chosen so that the sciencecraft remain close to relative rest at the vertices of an equilateral triangle with $\sim10^6$~km legs. We refer to the relative position of the sciencecraft as the \emph{constellation}. Tidal forces on the constellation owing to Sol, the Earth-Moon system, Jupiter and Venus all act to distort this configuration. The LISA mission lifetime is bounded by an upper limit on the relative sciencecraft velocities. This bound will be exceeded well-before deviations in the constellation shape affect significantly the angular sensitivity of the interferometer. Correspondingly, it is conventional to treat the constellation as a rigid equilateral triangle whose center follows a Kepler orbit, in the ecliptic plane, about Sol. We follow this convention. 

To evaluate LISA's response consider a (TT-gauge) plane gravitational wave propagating in the $\hat{\bm{n}}$ direction through Minkowski space:
\begin{align}
\bm{h} &= 
h_{+}(t-\hat{\bm{n}}\cdot\bm{x})\bm{e}_{+}(\hat{\bm{n}}) + 
h_{\times}(t-\hat{\bm{n}}\cdot\hat{\bm{x}})\bm{e}_{\times}(\hat{\bm{n}}), 
\end{align}
where $\bm{e}_{+}$ and $\bm{e}_{\times}$ are the usual $+$ and $\times$ wave polarization tensors,
\begin{subequations}
\begin{align}
0 &= {e}^{ij}_{+}n_j = e^{ij}_{\times}n_j\\
2 &= e_{+}^{ij}e_{+}^{kl}\delta_{ik}\delta_{jk} = e_{\times}^{ij}e_{\times}^{kl}\delta_{ik}\delta_{jk}\\
0 &= e_{+}^{jk}e_{+}^{jk}\delta_{ik}\delta_{jk}.
\end{align}
\end{subequations}
In the low-frequency (small antenna) limit the response of the inferometer may be expressed
\begin{subequations}
\begin{align}
r(t) & = 
\frac{L}{2}\left[
F_{+}(\bm{n},t)h_{+}(t-\bm{n}\cdot\bm{x}_c) + 
F_{\times}(\bm{n},t)h_{\times}(t-\bm{n}\cdot\bm{x}_c)\right]
\end{align}
where $L$ is the effective interferometer arm-length, $\bm{x}_c$ the constellation location, 
\begin{align}
F_{+} &= d_{ij}{e}_{+}^{ij}\label{eq:F+}\\
F_{\times} &= d_{ij}{e}_{\times}^{ij}\label{eq:Fx},
\end{align}
and $\bm{d}$ is the antenna projection tensor:
\begin{align}
\bm{d} &= \bm{u}\otimes\bm{u}-\bm{v}\otimes\bm{v},
\end{align}
\end{subequations}
with $\bm{u}$ and $\bm{v}$ the unit vectors in the direction of the interferometer arms. 

Our basic measure of LISA's sensitivity is the quadrature sum of $F_+$ and $F_{\times}$, 
\begin{align}
\frac{d\rho^2}{dt} &= F_{+}^2(\bm{n},t)+F_{\times}^2(\bm{n},t),
\end{align}
and its value integrated over a year-long observation and normalized so that its maximum value, over all $\bm{n}$, is unity: 
\begin{subequations}\label{eq:rho2}
\begin{align}
\widehat{\rho^2}(\bm{n}) &= \frac{\rho^2(\bm{n})}{\max_{\bm{n}'}\rho^2(\bm{n}')}
\end{align}
where
\begin{align}
\rho^2 &= \int_0^{1\,\mathrm{yr}}\frac{d\rho^2}{dt}dt.
\end{align}
\end{subequations}
Note that the dependence of $\rho^2$ on $\bm{n}$ is identical to the dependence on $\bm{n}$ of the power signal-to-noise associated with a compact object binary system when averaged over the binary's orientation.

\begin{figure}
\includegraphics[width=0.5\textwidth]{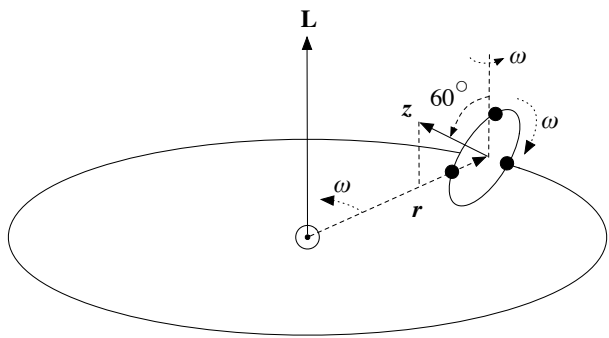}
\caption{Schematic of the LISA constellation in orbit about Sol. The constellation plane's normal is inclined $60\deg$ to the constellation's orbital angular momentum axis, is in the plane defined by the orbital radius vector and angular momentum, and is directed to have a positive projection on the orbital angular momentum axis. The constellation then ``orbits'' within its plane in a negative sense with the same period as the constellation's orbit about Sol: i.e., $\omega=2\pi~\text{yr}^{-1}$. In this schematic the sciencecraft ``above'' the orbital plane are always further from Sol than those below the plane; the orientation of the constellation plane could just as well be arranged so that the opposite were true.}\label{fig:orbitingConstellation}
\end{figure}

The time dependence of $F_{+}(\hat{\bm{n}},t)$ and $F_{\times}(\hat{\bm{n}},t)$ is determined by the motion of the constellation as it orbits Sol. The special orbits that allow the constellation to maintain a relatively stable equilateral triangle configuration enforce upon the constellation a characteristic internal motion and orientation with respect to the constellation's orbital plane. The orientation and motion is particularly simple when described in terms of the constellation plane's orientation relative to its orbital plane about Sol, and the orientation of the constellation in the constellation's plane. The sciencecraft orbits require that the constellation plane is always inclined $60\deg$ relative to the ecliptic plane (i.e., the constellation's orbital plane) with the constellation plane normal always in the plane defined by the constellation's orbital radius vector and its orbital plane normal: i.e., the constellation plane precesses, in a positive sense, about the ecliptic plane normal with a one-year period. In addition to this precession, the constellation itself spins within its plane, also with a one-year period. Choosing the constellation plane normal's direction to have a positive projection on the ecliptic plane normal the constellation spins in its plane with a negative sense. 
Figure \ref{fig:orbitingConstellation} shows these three different motions --- orbital, precessional, and spin --- schematically. 

\begin{figure}
\includegraphics[width=0.5\textwidth]{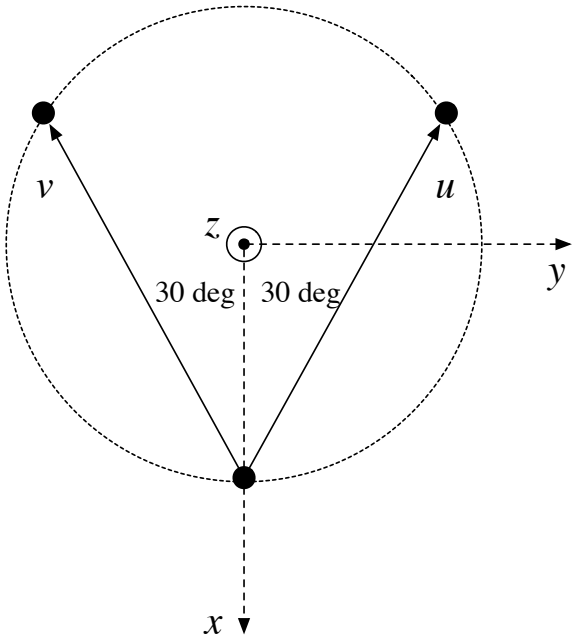}
\caption{The LISA constellation, gravitational wave antenna, and its rest-frame coordinates. The three sciencecraft form the vertices of an equilateral triangle. One sciencecraft acts as the beamsplitter of what is effectively a Michelson interfereomter; the other two sciencecraft act as the end stations. The ``beamsplitter'' sciencecraft is at the bottom of the figure. The interferometer arms, denoted $\bm{u}$ and $\bm{v}$ are shown as solid lines connecting the ``beamsplitter'' scienccraft to the ``end station'' sciencecraft. To define Cartesian rest-frame coordinates associated with the interferometer we identify the circle that circumscribes the three sciencecraft and choose the $x$-axis in the direction from the center of the circumscribed circle through the ``beamsplitter sciencecraft''. The $y$-axis is chosen in the constellation plane and orthogonal to the $x$-axis. The $z$-axis is chosen orthogonal to the plane so that three axes form a right-handed coordinate system. Note that this interferometer has four nulls --- directions toward which it is insensitive to any gravitational wave source --- in the $\pm\bm{x}$ and $\pm\bm{y}$ directions.}\label{fig:constellation}
\end{figure}

To evaluate $F_{+}(\hat{\bm{n}},t)$ and $F_{\times}(\hat{\bm{n}},t)$ via Equations (\ref{eq:F+})--(\ref{eq:Fx}) we need to express $\bm{e}_{+}$, $\bm{e}_{\times}$, $\bm{u}$ and $\bm{v}$ all in the same coordinate system. We choose to work in (heliocentric) ecliptic coordinates. To express $\bm{u}$ and $\bm{v}$, the unit vectors in the direction of the interferometer arms, in ecliptic coordinates we first describe them in the constellation's rest frame; next, we describe the rotations that relate ecliptic and constellation rest coordinates; and, finally, we express $\bm{u}$ and $\bm{v}$ in ecliptic coordinates via Equations \ref{eq:qRot}. Figure \ref{fig:constellation} shows the constellation, the interferometer arms, and the constellation rest-frame coordinates that we use to describe the constellation's gravitational wave response. In these coordinates $\bm{u}$ and $\bm{v}$ are
\begin{subequations}\label{eq:uvConstellation}
\begin{align}
\bm{u} &= -\bm{x}_c\cos\frac{\pi}{6} + \bm{y}_c\sin\frac{\pi}{6}\\
\bm{v} &= -\bm{x}_c\cos\frac{\pi}{6} - \bm{y}_c\sin\frac{\pi}{6}.
\end{align}
\end{subequations}

The convolution of the different rotations involved in describing the constellation's motion is quite simply expressed in terms of quaternions. Introduce the usual quaternion basis $\bm{i}$, $\bm{j}$ and $\bm{k}$, with
\begin{align}\label{eq:quaternions}
-1 &= \bm{i}^2 = \bm{j}^2 = \bm{k}^2 = \bm{ijk}.
\end{align}
Associate the quaternion basis vectors with the basis vectors of a Cartesian coordinate system: i.e., associate $\bm{x}$ with $\bm{i}$, $\bm{y}$ with $\bm{j}$ and $\bm{z}$ with $\bm{k}$. Any vector $\bm{n}$ may thus be regarded equally well as a quaternion or a Cartesian vector. It is then straightforward to show that the vector $\bm{n}'$ corresponding to the rotation of vector $\bm{n}$ by angle $\theta$ about axis $\bm{w}$ is equal to 
\begin{subequations}\label{eq:qRot}
\begin{align}
\bm{n}' &= \bm{q}_{w(\theta)} \bm{n} \bm{q}_{w(\theta)}^\dagger,
\end{align}
where
\begin{align}
\bm{q}_{w(\theta)} &= \cos\frac{\theta}{2} + \bm{w}\sin\frac{\theta}{2}\\
\bm{q}_{w(\theta)}^\dagger &= \bm{q}_{w(-\theta)}. 
\end{align}
\end{subequations}

To describe the relation between ecliptic and constellation rest-frame coordinates first introduce the usual ecliptic longitude and latitude $(l,b)$ and the corresponing Cartesian coordinate vectors $\hat{\bm{x}}_e$, $\hat{\bm{y}}_e$ and $\hat{\bm{z}}_e$. Measure the constellation's location in its orbit in terms of the position angle $\phi(t)=2\pi\omega t$, measured from $\bm{x}_e$, with $\phi=0$ at $t=0$. Finally, let $\phi_0$ denote the constellation plane orientation at $t=0$. The quaternion $\bm{q}_{ec}$ that relates vector components in the constellation coordinate system to components in the ecliptic coordinate system may be written
\begin{subequations}\label{eq:quaternion}
\begin{align}
\bm{q}_{ec} &= \bm{q}_3\bm{q}_2\bm{q}_1
\end{align}
where $\bm{q}_1$ orients the constellation in its plane,
\begin{align}
\bm{q}_1 &= \cos\frac{\phi_0-\phi}{2}+\bm{k}\sin\frac{\phi_0-\phi}{2};
\end{align}
$\bm{q}_2$ inclines the constellation plane relative to the ecliptic,
\begin{align}
\bm{q}_2 &= \cos\frac{\pi}{6}+\bm{j}\sin\frac{\pi}{6};
\end{align}
and $\bm{q}_3$ precesses the constellation plane about the ecliptic normal,
\begin{align}
\bm{q}_3 &= \cos\frac{\phi}{2}+\bm{k}\sin\frac{\phi}{2}.
\end{align}
\end{subequations}

It remains to express the polarization tensors $\bm{e}_{+}(\hat{\bm{n}})$ and $\bm{e}_{\times}(\hat{\bm{n}})$ in ecliptic coordinates. Let $l_n$ and $b_n$ identify the ecliptic longitude and latitude corresponding to the wave propagation direction $\bm{n}$ (i.e., the source is in the $-\bm{n}$ direction). A suitable choice of polarization tensors $\bm{e}_{+}$ and $\bm{e}_{\times}$ is then
\begin{subequations}
\begin{align}
\bm{e}_{+} &= {\bm{x}}_r\otimes{\bm{x}}_r-{\bm{y}}_r\otimes{\bm{y}}_r\\
\bm{e}_{\times} &= {\bm{x}}_r\otimes{\bm{y}}_r+{\bm{y}}_r\otimes{\bm{x}}_r
\end{align}
where
\begin{align}
{\bm{x}}_r &= \bm{q}_r {\bm{i}} \bm{q}_r^\dagger\\
{\bm{y}}_r &= \bm{q}_r {\bm{j}} \bm{q}_r^\dagger
\end{align}
for
\begin{align}
\bm{q}_r &= \bm{q}_2 \bm{q}_1\\
\bm{q}_2 &= \cos\frac{\pi/2-b_n}{2}+\bm{k}\sin\frac{\pi/2-b_n}{2}\\
\bm{q}_1 &= \cos\frac{l_n}{2}+\bm{j}\sin\frac{l_n}{2}. 
\end{align}
\end{subequations}

\section{Discussion}\label{sec:disc}

\begin{figure}
\includegraphics[width=0.5\textwidth]{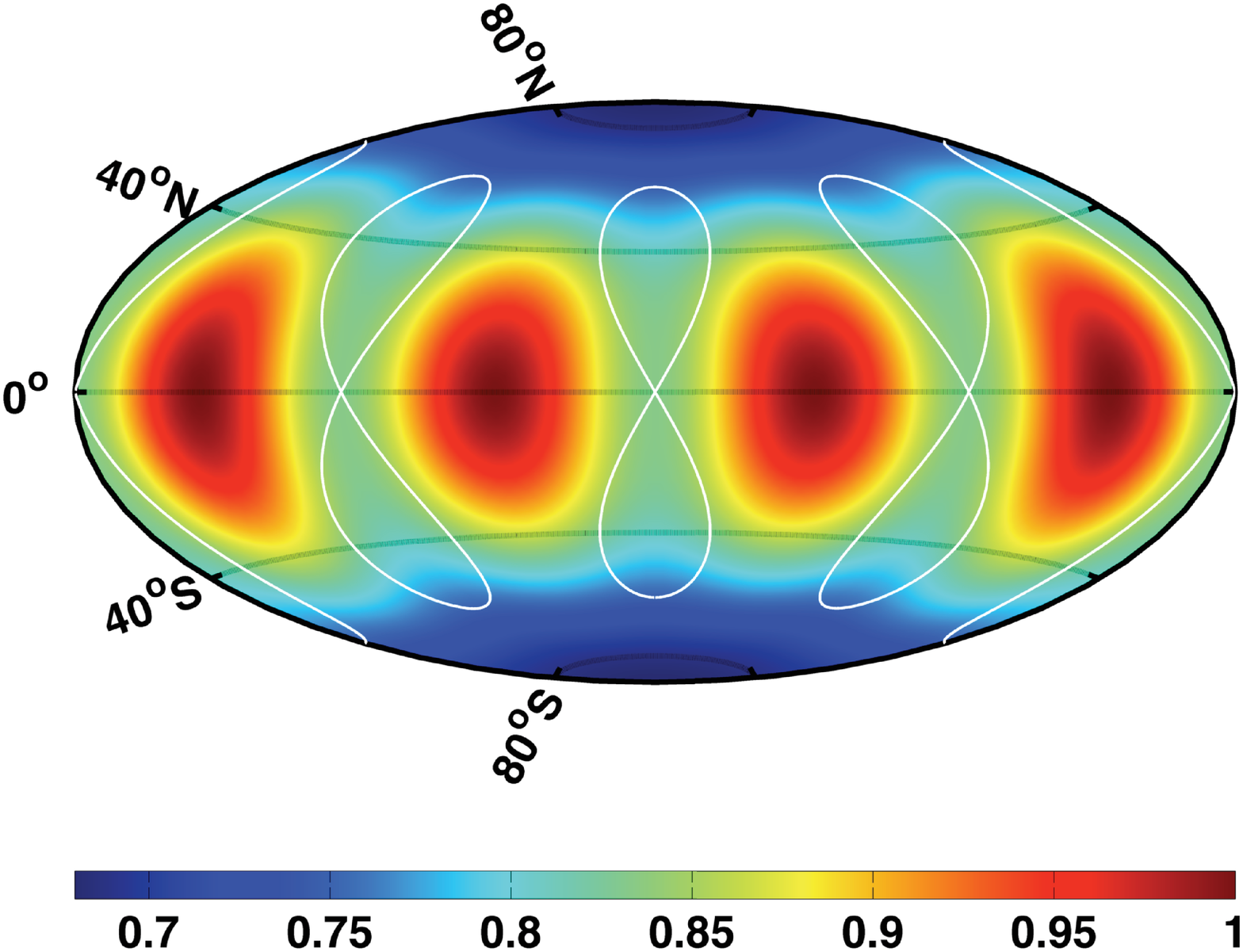}
\caption{Color map showing the (low-frequency) sensitivity $\widehat{\rho^2}$ as a function of the wave propagation direction. The white lines trace out the paths of the antenna nulls over an annual period. Ecliptic longitude is unmarked: its calibration depends on the initial orientation of the constellation.}\label{fig:8s}
\end{figure}

An interferometric detector has a null response to gravitational-wave sources in four different directions --- all in the plane formed by the detector. When the light travel time along its arms is equal, as is the case for LISA, two of the nulls are in the direction and anti-direction of the bisector between the arms, while the other two nulls are at right angles to these directions: i.e., in the $\pm\bm{x}$ and $\pm\bm{y}$ directions in Figure \ref{fig:constellation}. As the antenna orbits Sol these nulls trace out figure-8 patterns on the sky. The white curves in Figure~\ref{fig:8s} show, in ecliptic coordinates, the trace of these nulls over a full orbit overlaid on a colormap that shows our sensitivity measure $\widehat{\rho^2}$ (see Eq.~\ref{eq:rho2}). 

Note that the ecliptic longitude is unlabeled in Figure~\ref{fig:8s}. The ecliptic longitude of the traces of the nulls depend on the constellation's initial orientation $\phi_0$ (see Eq.~\ref{eq:quaternion}) at $t=0$: i.e., rotating the constellation in its plane while leaving the plane's location and orientation fixed shifts the null curves in ecliptic longitude by the same angle. It is the initial choice of constellation orientation $\phi_0$ that allows us to ``point'' LISA toward or away from any source that lies within $60\deg$ of the ecliptic plane. 

Although all but the closest extragalactic sources of gravitational radiation are likely to be distributed isotropically, Galactic sources --- including the vast majority of close white dwarf binaries, by far the largest expected population of LISA sources --- will be concentrated primarily in the direction of the Galactic Center and, secondarily, the Galactic plane. \emph{It is a fortunate coincidence that the Galactic Center lies nearly on the ecliptic plane and the Galactic Plane, like the LISA constellation plane, is inclined $60\deg$ relative to the ecliptic plane:} just those regions on the sky where the choice of $\phi_0$ gives us greatest control over LISA's sensitivity. Referring to Figure \ref{fig:8s}, our initial choice of LISA constellation orienation can make a 17\% difference in the constellation's sensitivity toward the Galactic Center sources: i.e., to just those sources that dominate LISA's signal. If we choose as our goal the study of these sources we can maximize LISA's sensitivity to the Galactic Center direction; alternatively, if our goal is the detection and study of extragalactic sources we can choose LISA's initial orientation to minimize our sensitivity to these (now confounding) sources, allowing LISA observations to ``dig deeper'' into the pool of extra-galactic sources. 

\begin{figure}
\includegraphics[width=0.5\textwidth]{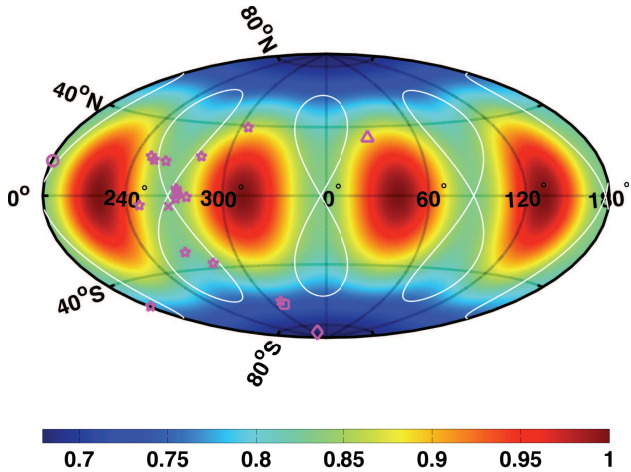}
\caption{The locations of the objects listed in Table~\ref{tab:sourceloc} plotted on top of the LISA sensitivity when the constellation's initial orientation is chosen to minimize its sensitivity, as measured by the sensitivity measure $\widehat{\rho^2}$ (see Eq.~\ref{eq:rho2}), to Galactic Center sources. The Galactic Center is marked by an $\times$, globular clusters are marked by stars, M31 is marked by a triangle, the LMC by a square and the SMC by a diamond, and the Virgo cluster by an open circle.}\label{fig:obj}
\end{figure}

Now consider the binary choice of maximizing or minimizing LISA sensitivity to Galactic Center sources and ask what effect this choice has on the sensitivity to gravitational wave emitters associated with other localized and nearby objects: in particular, globular clusters within 5 kpc of Earth, the Large and Small Magellanic clouds, the Andromeda galaxy (M31), and the Virgo cluster. These objects are representative of typical hosts for halo and extragalactic populations of white-dwarf binaries, ultra compact binaries, and nearby extreme mass-ratio inspiral systems. Figure~\ref{fig:obj} and Table~\ref{tab:sourceloc} summarize how minimizing or maximizing sensitivity to Galactic Center sources affects LISA's sensitivity to gravitational wave emitters associated with these objects. Figure~\ref{fig:obj} plots the object locations, given in the first two columns of Table~\ref{tab:sourceloc}, together with the antenna nulls and sensitivity measure corresponding to an initial LISA orientation that suppresses the constellation's response to Galactic Center sources. Both Magellanic clouds (diamond for the LMC, square for the SMC) and two globular clusters (47 Tuc and NGC 3201, both marked with stars) are sufficiently far south of the ecliptic that there is negligible change in the sensitivity for any constellation orientation. Most of the remaining globular clusters (all marked with stars) are close enough to the Galactic Center that LISA's sensitivity to sources within them varies together with sensitivity to the Galactic Center. Only for the globular cluster M71 and the Andromeda galaxy (M31) is the constellations's sensitivity relatively high when it is relatively low for Galactic Center sources. The final column of Table~\ref{tab:sourceloc} quantifies these observations, providing for each object the ratio $\widehat{\rho^2}_{\min}/\widehat{\rho^2}_{\max}$, where $\widehat{\rho^2}_{\max}$ ($\widehat{\rho^2}_{\min}$) is the sensitivity measure when the constellation's orientation is chosen to (maximize (minimize) its sensitivty to Galactic Center sources. 

\begin{table}
\caption{Pointing LISA to minimize or maximize its sensitivity to Galactic Center sources affects its sensitivity to gravitational wave sources in other objects. Here is shown the ratio $\widehat{\rho^2}_{\min}/\widehat{\rho^2}_{\max}$, where $\widehat{\rho^2}_{\min}$ ($\widehat{\rho^2}_{\max}$) is the sensitivity measure for the corresponding source directions when LISA's initial orientation is chosen to minimize (maximize) its sensitivity to Galactic Center sources. The Sources column abbreviations are DWD for double white-dwarf binaries, COB for compact object (neutron star or stellar mass black hole) binaries, and EMRI for extreme mass-ratio inspiral systems.}\label{tab:sourceloc}
\begin{tabular}{l|rrlr}
Object&{Lon}&{Lat}&{Sources}&$\widehat{\rho^2}_{\min}/\widehat{\rho^2}_{\max}$\\
\hline
 Gal. Cen. & 266.85172  &    -5.6076736 & DWD, COB & 0.84 \\
 Virgo Cluster &  181.04266  &     14.333893 & COB, EMRI & 0.82\\
 LMC   & 312.50989   &   -85.351425 &   COB & 1.00\\
 SMC   &  312.08823   &   -64.605469   & COB & 0.99\\
 Androm. Gal. (M31)& 27.849274  &     33.349022  & COB, EMRI & 1.04\\
 NGC104 (47Tuc) &  311.25247 &     -62.352768    & DWD, COB & 1.03\\
 NGC3201    &181.36612   &   -51.539581    & DWD, COB  & 0.97\\
 NGC6121 (M4) &  248.48676 &     -4.8687668    & DWD, COB & 0.96\\
 NGC6218 (M12) & 250.57292  &     20.272308 & DWD, COB & 0.97\\
 NGC6259 (M10) & 253.45894  &     18.441391  & DWD, COB &  0.94\\
 NGC6366   & 261.54391  &     18.122971    & DWD, COB  & 0.88\\
 NGC6397  & 266.69220  &    -30.290436  & DWD, COB & 0.90\\
 NGC6544  & 271.66437  &    -1.5686417    & DWD, COB & 0.83 \\
 2MS\_GC01  & 271.97073  &     3.5952682    & DWD, COB & 0.83 \\
 2MS\_GC02 &   272.24817 &      2.6416550    & DWD, COB & 0.83 \\
 Terzan12  &  272.82751  &    0.66725922    & DWD, COB & 0.84 \\
 NGC6656 (M22) &  278.31403  &   -0.72771454  & DWD, COB & 0.86\\
 GLIMPSE01 &  283.12085  &     21.387779    & DWD, COB & 0.91\\
 NGC6752   & 281.02106  &    -37.221313   & DWD, COB & 0.95\\
 NGC6838 (M71)  &  305.34909   &    38.792225    & DWD, COB & 1.05
\end{tabular}
\end{table}

\begin{figure}
\includegraphics[width=0.5\textwidth]{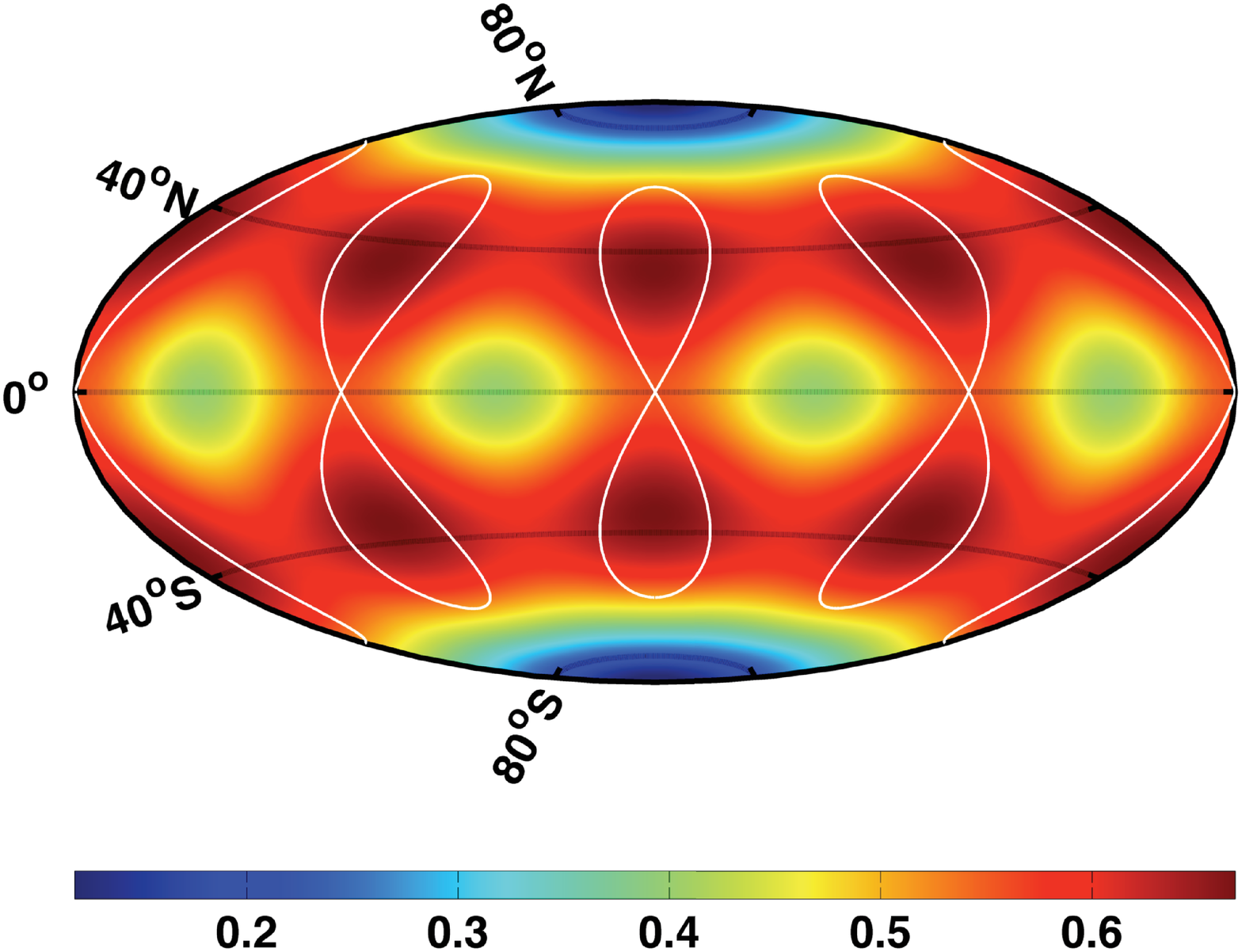}
\caption{The fractional root-mean-square variation of $d\rho^2/dt$ from its mean value over an orbital period. The white lines trace out the paths of the nulls.}\label{fig:var}
\end{figure}

Now turn attention to the variation in LISA sensitivity to fixed sources over the course of a year. As the constellation rotates in its plane $d\rho^2/dt$ is modulated for all sources but those in the direction of the ecliptic poles. For sources within $60\deg$ of the ecliptic plane there will be choices of orientation that lead an antenna null to pass over the source at least once per year, corresponding to 100\% modulation of antenna's response to the incident waves. Figure~\ref{fig:var} summarizes the modulation of the response as the fractional root-mean-square variation of $d\rho^2/dt$ from its mean value over an orbital period: i.e., 
\begin{align}
\sigma(\bm{n}) &= \rho^{-}\left[\int_0^{1\,\text{yr}}\left[(d\rho^2/dt)^2-\rho^2\right]dt\right]^{1/2}. 
\end{align}
For sources at ecliptic latitudes $35\deg\lesssim |b| \lesssim 60\deg$ the modulation is greater than $\sim65$\% regardless of the initial constellation orientation. Referring to Table~\ref{tab:sourceloc} this includes three of the nearest globular clusters (NGC3201, NGC6752 and NGC6838). For sources with ecliptic latitude $|b|\lesssim35\deg$ the degree of modulation varies significantly with the initial constellation orientation: for sources on the ecliptic equator it can be varied between $40\%$ and $65$\%.Only for sources near the ecliptic poles ($|b|>80\deg$) will the modulation ever be small ($\lesssim30$\%). 

The modulation of LISA's sensitivity in a given direction provides another means of ``tuning'' LISA's sensitivity to different sources. 
It has often been remarked that the number of low-frequency sources in the direction of the Galactic Center is sufficiently great that their cacophony may limit our ability to observe other low-frequency sources elsewhere on the sky \citep{bender:1996:lli,kosenko:1998:ogw}. Taking advantage of our ability to point LISA by choosing its initial orientation, we can arrange for the antenna nulls pass over or close to the Galactic Center twice per year. During these periods LISA will largely insensitive to Galactic Center sources while remaining sensitive to sources at moderate to high ecliptic latitudes, and low ecliptic latitudes at longitudes $\pm45\deg$ and $\pm135\deg$ from the Galactic Center. This includes most of the sources identified in Table~\ref{tab:sourceloc}. 

\begin{figure}
\includegraphics[width=0.5\textwidth]{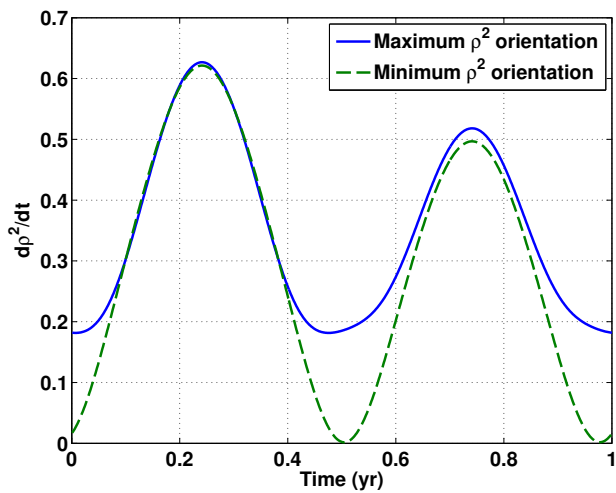}
\caption{The instantaneous contribution to the sensitivity measure for sources in the direction of the Galactic Center when the constellation's orientation is chosen to maximize or minimize its annual sensitivity to Galactic Center direction sources. Note how the contribution to the sensitivity measure can be made to vanish twice per orbit.}\label{fig:gcMinMax}
\end{figure}

\begin{figure}
\includegraphics[width=0.5\textwidth]{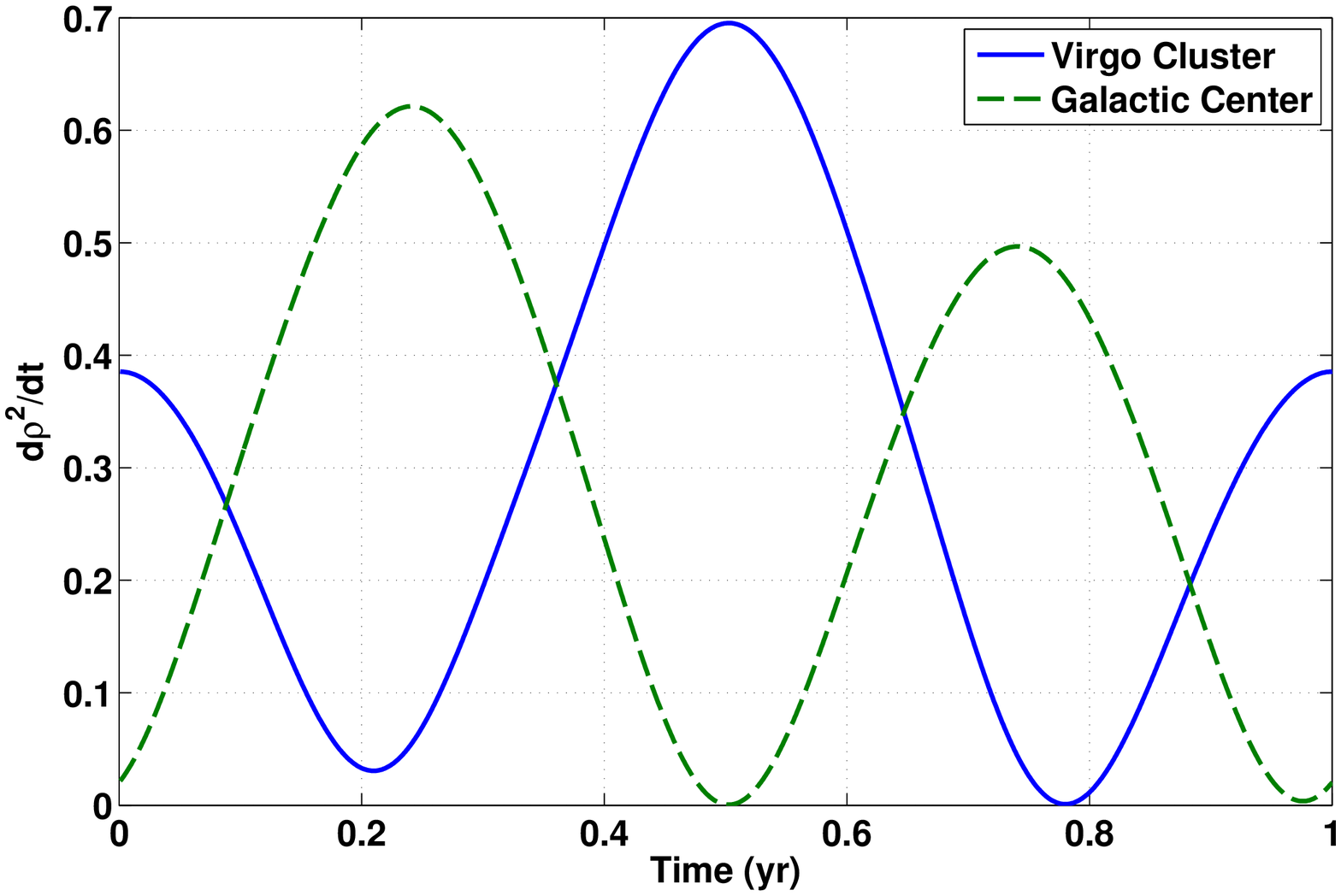}
\caption{The instantaneous contribution to the sensitivity measure for sources in the direction of the Virgo Cluster and the Galactic Center when the constellation's orientation is chosen to minimize its annual sensitivity to Galactic Center direction sources. Note how for this  orientation the constellation's sensitivity to Galactic Center sources vanishes when its sensitivity to Virgo Cluster sources is greatest, and is large only when the sensitivity to Virgo Cluster sources is small. By appropriate pointing of the LISA constellation the interference of Galactic Center sources in the detection of Virgo Center sources can be minimized.}\label{fig:vcgc}
\end{figure}

To illustrate, Figure~\ref{fig:gcMinMax} shows how the modulation in sensitivity varies over the course of a year for Galactic Center direction sources when LISA's initial orientation is chosen to maximize or minimize its sensitivity in this direction. While the peak sensitivity over the course of a year varies only slightly, the minimium sensitivity can be sent to zero. 
Figure~\ref{fig:vcgc} shows LISA's sensitivity to sources in the direction of the Virgo Cluster and the Galactic Center when LISA's orientation is chosen to so that an antenna null passes over the Galactic Center twice per year. \emph{For this orientation LISA's sensitivity to sources in the direction of the Galactic Center vanishes just when its sensitivity to sources in the direction of the Virgo Cluster is greatest, and vice versa.} Thus, by appropriate pointing of LISA we can mitigate significantly any interference that Galactic close white dwarf binaries have in the detection of extra-galactic sources. 

\section{Conclusions} \label{sec:concl}

The LISA sciencecraft form what is effectively an interferometric gravitational wave antenna. Such an antenna has four nulls: i.e., directions in which it is entirely insensitive to incident gravitational waves. As the sciencecraft proceed in their special orbits the antenna rotates and precesses, leading the nulls to trace-out four figure-8 pattens on the sky. The choice of initial constellation orientation can be chosen to ensure that a null traces over, or avoids, any fixed sky location within $60\deg$ of the ecliptic plane. Sources in this zone include the Galactic Center, the Virgo Cluster, and the nearest Galactic globular clusters. By choice of the initial constellation orientation LISA may thus be pointed toward or away from any of these sources, with significant consequences for the success of its science mission. In the case of sources in the direction of the Galactic Center this freedom amounts to a 17\% difference in the power signal-to-noise accumulated over a year's observation (averaged over all source orientations). 

As important as the variation in the annual signal-to-noise is the variation in the sensitivity \emph{throughout} the sciencecraft constellation's annual orbital period. The relative location of the Galactic Center and the Virgo Cluster is such that we may choose a null to cover the Galactic Center during just those times when the antenna's sensitivity to the Virgo Cluster is greatest. Thus, we can arrange our orbits to suppress the confusion noise associated with the large number of Galactic white dwarf binaries at just those times when our sensitivity to Virgo Cluster sources is greatest. 

The choice of initial constellation orientation need not affect the mission cost. The dominant cost to the mission associated with a choice of initial constellation  orientation is that required to launch the mass in fuel needed to move the sciencecraft from Earth to their initial orbital stations. Viewed with respect to Earth there is a constellation orientation that requires minimum fuel mass to reach. By appropriate choice of launch date we can make this cost-preferred Earth-relative configuration correspond to any desired ecliptic coordinate relative orientation. 

How to best take advantage of the ability to point a LISA-like gravitational wave antenna requires the kind of judgements that are only possible in the context of the mission science goals: e.g., is the mission's principal science goal the study of Galactic sources, extra-galactic sources, or something different still? Here we have demonstrated that the mission profile includes a heretofore unrealized freedom --- the ability to point the antenna toward or away from sources within $60\deg$ of the ecliptic plane --- and have given but one example of how choices in pointing can affect sensitivity to multiple sources.

\acknowledgments
MJB and LSF acknowledge the hospitality of the Aspen Center for Physics, supported in part by the National Science Foundation under Grant No. 1066293. KJ and LSF acknowledge the support of NSF awards 0940924 and 0969857. MJB acknowledges the support of NASA grant NNX08AB74G and the Center for Gravitational Wave Astronomy, supported by NSF Cooperative Agreement HRD-0734800 and HRD-1242090. 


\end{document}